\documentclass[%
 reprint,
 amsmath,amssymb,
 aps,
]{revtex4-2}

\usepackage{graphicx}
\usepackage{dcolumn}
\usepackage{bm}
\usepackage{color}

\begin{document}

\preprint{APS/123-QED}

\title{Effects of Polydispersity on the Plastic Behaviors of Dense 2D Granular Systems Under Shear}
\author{Yonglun Jiang}
\affiliation{Department of Physics, Emory University,
Atlanta, GA 30322, USA}
\email{yjia249@emory.edu, erweeks@emory.edu}

\author{Daniel M. Sussman}
\affiliation{Department of Physics, Emory University,
Atlanta, GA 30322, USA}

\author{Eric R. Weeks$^*$}
\affiliation{Department of Physics, Emory University,
Atlanta, GA 30322, USA}

\date{\today}

\begin{abstract}
We study particle-scale motion in sheared highly polydisperse amorphous materials, in which the largest particles are as much as ten times the size of the smallest. We find strikingly different behavior from the more commonly studied amorphous systems with low polydispersity.  In particular, analysis of the nonaffine motion of particles reveals qualitative differences between large and small particles: the smaller particles have dramatically more nonaffine motion, which is induced by the presence of the large particles.  We characterize the crossover in nonaffine motion from the low- to high-polydispersity regime, and demonstrate a quantitative way to distinguish between ``large'' and ``small'' particles in systems with broad distributions of particle sizes.

\end{abstract}

\maketitle


\label{intro}

Amorphous materials are common, ranging from glasses to emulsions, foams, granular media, cement pastes, food products, and more.  These materials are often composed of mixtures of various sizes of particles.  Much prior work has studied the flow of amorphous materials using model systems with low polydispersity, which is to say, mixtures of particles of fairly similar sizes \cite{liu96,yamamoto97,mason97emulsions,hebraud97,falk98,losert00,petekidis02,teitel07,schall07,utter08,lemaitre09,chen10,manning11,cubuk15,patinet2016connecting,vasisht18,hassani19,tsai21}.  However, many natural materials are highly polydisperse, with particle sizes varying by factors of ten or more, which impacts the flow of glaciers \cite{haeberli06}, landslides and avalanches \cite{pitman05}, soil \cite{or02}, mud \cite{besq03}, cement \cite{rosquoet03}, and food products \cite{taylor09}. Computational studies have also increasingly relied upon highly polydisperse model glassformers, where manipulating particle sizes has allowed studies of equilibrium glass configurations at previously unprecendentedly low temperatures~\cite{ninarello2017models,berthier2017configurational,brito2018theory,kapteijns2019fast}. These natural and model systems with large polydispersity are complex and spatially heterogeneous, and it is hard to extract general principles.  The goal of this Letter is to examining the role of the particle size distribution in sheared materials and, in so doing, bridge between simple model systems with low polydispersity and complex highly polydisperse real-world materials.

Polydispersity leads to interesting physics.  For example, polydisperse hard spheres can phase separate into multiple crystalline phases \cite{sollich10}.  Polydispersity can lead to new phases for active matter systems \cite{kumar21}.  An experimental study of polydisperse colloidal glasses found that different particle sizes had different dynamics and local environments \cite{heckendorf17}.  Diffusion of tracers in porous materials becomes anomalous when the porous medium is highly polydisperse \cite{cho12}.  Force chains in granular materials become dramatically more heterogeneous in more polydisperse systems \cite{nguyen14,nguyen15,cantor18}.  The viscosity of particulate suspensions strongly depends on polydispersity \cite{pednekar18}, varying by orders of magnitude for constant volume fraction of particles \cite{chong71}.  These studies highlight the role of the particle size distribution in leading to new physics.  In contrast, prior work studying sheared amorphous materials typically downplays the role of particle size distributions.  Experiments and simulations of sheared amorphous materials often use slightly polydisperse samples to inhibit the formation of crystalline structures:  for example, two distinct sizes with size ratio $O(1)$ \cite{yamamoto97,falk98,teitel07,utter08,lemaitre09,manning11,cubuk15,hassani19} or experiments using nominally single component systems with small intrinsic polydispersity \cite{liu96,mason97emulsions,hebraud97,petekidis02,schall07,chen10,vasisht18,tsai21}.  These studies have led to insights such as the importance of non-affine motion in sheared disordered materials \cite{falk98,utter08,schall07}, but generally treat the amorphous system as homogeneous.  Exceptions to this treatment exist; for example some important studies show that ``soft spots'' in amorphous materials are more likely to exhibit particle rearrangements under shear \cite{manning11}, although even in these analyses, it is common to focus on identifying soft spots centered only on larger particles in a bidisperse mixture \cite{cubuk15,schoenholz16,bapst2020unveiling}. It is far from clear that many of the methods used to identify these disordered ``defects'' in the solid will generalize to highly polydisperse samples, or that particles of different sizes within a highly polydisperse sample even qualitatively show similar non-affine behavior under shear.  Indeed, a confocal microscopy study of a sheared highly polydisperse emulsion showed qualitative differences in the motion of large and small droplets \cite{clararahola15}, although it did not vary the particle size distribution.

In this Letter we show that sheared highly polydisperse 2D amorphous systems are qualitatively different from systems of low polydispersity. We show that large particles behave qualitatively differently from small particles; we demonstrate how to quantify which particles are ``large'' and ``small;'' and we show how these effects appear as the particle size distribution broadens.  Additionally, our results show that the largest particles strongly influence nearby particles to rearrange, suggesting that previously studied soft spots (e.g. \cite{manning11,cubuk15,schoenholz16,bapst2020unveiling,boattini2021averaging}) will be different in character -- and in some cases easier to identify -- in highly polydisperse materials.

\begin{figure}[h]
    \centering
    \includegraphics[width=8.6cm,bb= 20 120 580 350]{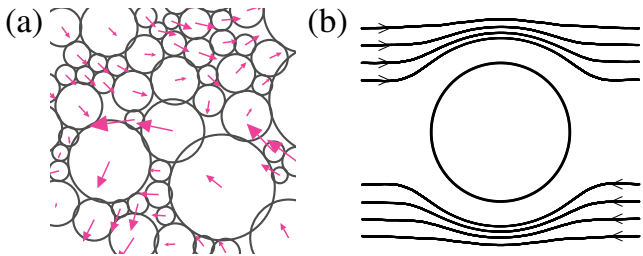}
    \caption{(Color online). Panel (a) shows a snapshot of a system with an exponential size distribution, with the particle size ratio $R_{\max}/R_{\rm min}= \alpha = 10$. Pink arrows indicate the motions of particles for a strain interval of $0.005$.  Panel (b) sketches the mean flow pattern around large particles under the applied shear strain.  
    }
    \label{fig:sample_example}
    \vspace*{-12pt}
\end{figure}

In our two-dimensional simulations we use the Durian bubble model \cite{durian95}, where particles experience viscous forces from neighboring particles moving at different velocities and repulsive contact forces. We consider a variety of truncated exponential size distributions, $P(R) \sim \exp(-R/\lambda)$ where $R$ is the radius, considered over the domain $R_{\rm min} \leq R \leq R_{\rm max} \equiv \alpha R_{\rm min}$. We study systems in which $\alpha$, which characterizes the width of the distribution, ranges from 2 to 10.  The decay constant $\lambda$ is set to $R_{\rm min}$.  We take $\langle R \rangle$ as our unit of length, and we nondimensionalize time by the microscopic relaxation time (based on the inter-particle spring constant and viscous damping forces \cite{durian95}). We will compare the properties  of these exponentially distributed systems with a standard bidisperse mixture composed of equal numbers of particles whose size ratio is $1 : 1.4$. All of these distributions can be characterized by their polydispersity, defined as the standard deviation of $P(R)$ divided by $\langle R \rangle$.  The polydispersity ranges from 0.20 to 0.50 for the exponentially decaying $P(R)$ we consider, and is 0.17 for the bidisperse sample.  We shear these systems in square boxes with length $L$ using Lees-Edwards boundary conditions. We keep $\frac{L}{\langle R \rangle} = 100$ constant which guarantees that $L$ is at least $20R_{\rm max}$  for all systems. The area fraction $\phi$ is 0.93, which is well above the jamming transition point ($\phi_c= 0.84 - 0.86$ for our particle size distributions).  Our nondimensional strain rate is $\dot{\gamma} = 10^{-4}$, chosen to be in a rate-independent regime \cite{durian95,ono02,lemaitre09}. We simulate the shear at least up to strain $\gamma=10$ to ensure enough statistics; an initial transient response for $\gamma < 0.2$ is discarded before analysis.  Other simulation details are discussed in Ref.~\cite{hong17}. We will focus most of our discussion on the $\alpha=10$ (maximally polydisperse) systems of $2500$ particles and the bidisperse systems of $3098$ particles unless otherwise stated. A snapshot of a portion of the polydisperse system under shear is shown in Fig.~\ref{fig:sample_example}(a).

We examine particle motion over small strain intervals $\Delta \gamma = 0.005$, an interval over which particles do not rearrange dramatically. To characterize the behavior of individual particles, we consider the non-affine component of motion by subtracting off the mean (affine) flow:
\begin{equation}\label{eqnnonaffine}
    \Delta \vec{r}_{{\rm NA},i} = \Delta \vec{r}_{i} - \Delta \gamma \, y_{i} \cdot \hat{x},
\end{equation}
where for particle $i$, the first term on the right hand side is the real motion over the strain interval, and the second term on the right hand side is the affine motion imposed by the simulation, where $\hat{x}$ is the velocity direction and $y_{i}$ is the position of particle $i$ in the gradient direction. Local rearrangements cause deviations from purely affine motion, as seen in Fig.~\ref{fig:sample_example}(a):  were the motion entirely affine, all arrows would be horizontal. We note that our conclusions are unaffected by the precise choice of measure of non-affinity of particle motion. For instance, using $D^2_{\rm min}$ (as introduced by Falk and Langer \cite{falk98}), to quantify the non-affine motion of groups of particles gives qualitatively the same results as considering the statistics of this single-particle measure of non-affinity.

\begin{figure}[h]
    \centering
    \includegraphics[width=8cm]{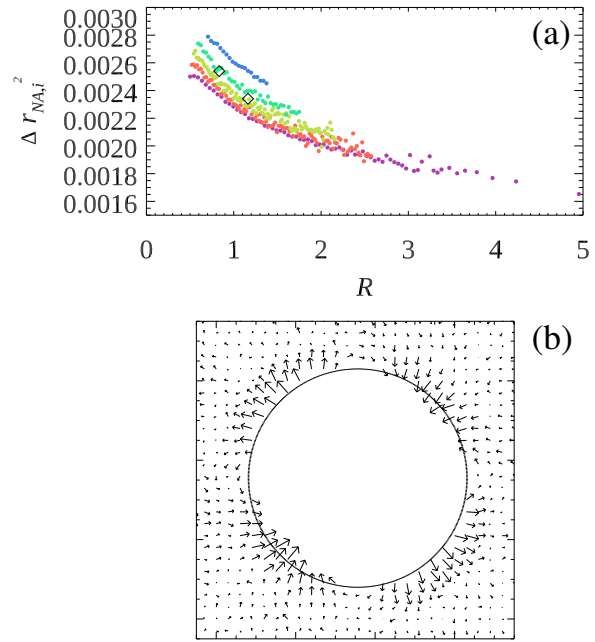}
    \caption{(Color online). (a) Averaged $\Delta \vec{r}_{\rm NA,i}^{\,2}$ versus particle radius $R$ for systems with different size distributions.  From top curve (blue) to bottom (purple) the symbols symbols correspond $\alpha = 2, 3, 4, 5,$ and $10$.  The two black diamonds are for the bidisperse particle size distribution. (b) Measured nonaffine motion field $\langle \Delta r_{\rm NA} \rangle(x,y)$ around reference particles with $2 \leq R_i \leq 2.8$ in the system with $\alpha=10$.}
    \label{fig:ndr_radius}
    \vspace*{-12pt}
\end{figure}

To understand how nonaffine motion depends on particle size, we calculate the mean $\Delta \vec{r}_{\rm NA,i}^{\,2}$ as a function of particle size $R$, shown in Fig.~\ref{fig:ndr_radius}(a).  On average $\Delta \vec{r}_{\rm NA,i}^{\,2}$ decreases with $R$ for all systems, including the bidisperse system; this agrees qualitatively with previous observations in polydisperse emulsions under cyclic shear \cite{clararahola15}.  Figure \ref{fig:ndr_radius}(a) shows that large particles are more likely to follow the affine shear flow, whereas small particles will have more shear-induced diffusivity.  A simple explanation is that large particles have more neighbors than small ones. The influence of these neighbors on the motion of the large particles on average cancel with each other, which results in the larger particles having smaller magnitude of $\Delta \vec{r}_{\rm NA,i}$.  Equivalently, moving a large particle non-affinely requires more neighboring particles to also move non-affinely to make room, which is harder to do.  The data from our $\alpha=10$ systems in Fig.~\ref{fig:ndr_radius}(a) are well fit by $\Delta \vec{r}_{\rm NA,i}^{\, 2} \sim {R}^\beta$ with $\beta = -0.18$, although the data do not span a big enough range to conclusively decide that this is power law behavior.  Note that the bidisperse results also match to the family of curves, showing measurably different $\Delta \vec{r}_{\rm NA,i}^{\,2}$ values for small and large particles.

These results reveal the following microscopic picture of motion near the large particles.  Large particles are ``strong'' and have less nonaffine motion; they are more likely to follow the affine imposed shear flow.  In the reference frame co-moving with the affine velocity of a large particle, this relative immobility causes the ``weaker'' small particles to detour around the larger particles, as sketched in Fig.\ref{fig:sample_example}(b).  Indeed, it is this detour motion that gives the smaller particles their larger average $\Delta \vec{r}_{\rm NA,i}^{2}$ seen in Fig.~\ref{fig:ndr_radius}(a).  Examining trajectories of individual particles reveals motions that qualitatively match the sketch of Fig.~\ref{fig:sample_example}(b) (data not shown).

To better understand how large particles perturb the flow we calculate the average non-affine flow field around particles of different sizes.  To do this, we average the non-affine motion $\langle \Delta \vec{r}_{{\rm NA},j} \rangle$ of all particles $j$ at a specific position $(x,y)$ relative to a reference particle $i$.  We then average that field over all reference particles $i$ with radii $R_i$ in a specific range to get better statistics yielding $\langle \Delta \vec{r}_{{\rm NA}}\rangle(x,y)$.  In Fig.~\ref{fig:ndr_radius}(b) we show this average field for particles with $2.0 \leq {R}_i \leq 2.8$.  At the top left and bottom right the mean non-affine flow field is outward, whereas at the top right and bottom left the mean non-affine flow is inward.  The top left and bottom right, relative to the reference particle, are referred to as the ``compressive directions'' as the imposed affine flow tries to push neighboring particles toward the reference particles \cite{batchelor_determination_1972,bergenholtz02}.  This affine push is resisted by the large reference particle, resulting in outward-pointing non-affine motion.  Likewise, the regions at the top right and bottom left are referred to as the ``extensional directions'' in terms of the background flow, and the non-affine motion is inward.  Adding the background affine shear flow to the nonaffine flow field of Fig.~\ref{fig:ndr_radius}(b) yields the qualitative sketch of Fig.~\ref{fig:sample_example}(b).  This non-affine motion field clearly illustrates the importance of relative positions in the polydisperse sample.

\begin{figure}
    \centering
    \includegraphics[width=8cm]{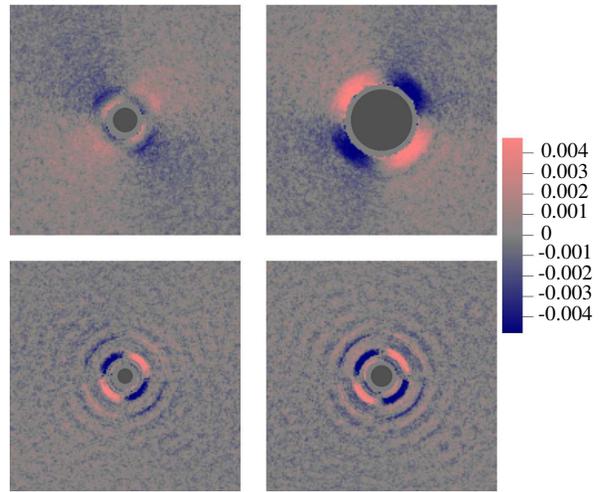}
    \caption{Color field of $\Delta \vec{r}_{\rm NA,i} \cdot \hat{r}$; the dot product with $\hat{r}$ selects for components of the motion that are outward (light red) or inward (dark blue), as indicated by the color bar. From left to right, the top two panels are size ranges $R_i = 0.80-0.84$ and $2.0-2.8$ using data from the broadest size distribution ($R_{\rm max}/R_{\rm min}=\alpha=10$).  The bottom two panels are from the bidisperse system (particles in size ratio $1:1.4$) for the small (lower left) and large (lower right) particles.}
    \label{fig:ndr_field}
    \vspace*{-20pt}
\end{figure}

How do the flow patterns measured by $\Delta \vec{r}_{\rm NA}(x,y)$ depend on the reference particle size? Figure \ref{fig:ndr_field} shows four examples of the $\hat{r}$ component of this field.  The top two panels are data from the broadest particle size distribution, examining the flow around smaller (top left) and larger (top right) particles.  For comparison, the bottom two panels are from the simulation with the bidisperse distribution, again showing the smaller (bottom left) and larger (bottom right) of the two particle sizes.  We first highlight that sign of the field in the top left panel is opposite to the field in the top right panel. We also note that in this system a relatively small but still finite far-field can be identified. We also highlight that this behavior is qualitatively different to the data shown for the bidisperse samples in the bottom panels: there the smaller and larger particles have qualitatively similar non-affine flow fields, with rings related to the pair correlation function and a vanishing far field.

The interpretation of the top panels of Fig.~\ref{fig:ndr_field} is that large particles are strong, move more affinely, and force the other particles to detour around them.  For the smaller reference particles, the influence of the reference particle is clearly different.  At the surface of these small reference particles, for a center of a neighboring particle to be close, the neighboring particle must also be small.  Thus the region immediately around the small reference particle looks similar to the region around the large particle:  small reference particles cause an outward non-affine motion along the compressive direction, and inward non-affine motion along the extensional direction.  However, farther away from small reference particles, the size of neighboring particles can be significantly larger than the reference particle.  These small reference particles are weaker and more likely to be moved nonaffinely by their neighboring particles.  Thus, the inward moving (dark blue) colors around the small reference particle along the compressive directions reflect that, on average, the small reference particle is being pushed away from the neighbors along these directions.  In other words, the non-affine motion pattern around large reference particles, as seen in Fig.~\ref{fig:ndr_field}(top right), is precisely because the large reference particles are larger than many other particles; and the pattern around smaller reference particles is qualitatively different precisely because they are smaller than many other particles.

\begin{figure}
    \centering
    \includegraphics[width=8cm]{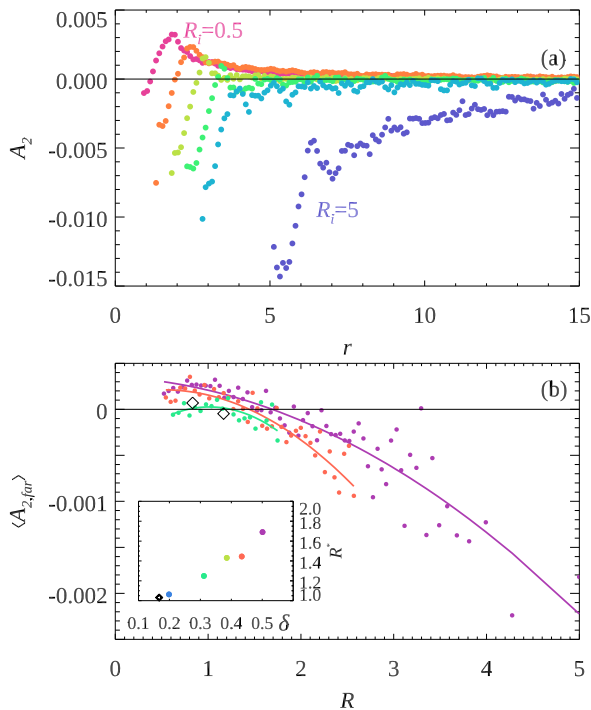}
    \caption{(a) Prefactor characterizing the non-affine field, $A_2$, versus distance, $r$, for several reference particle sizes in the $\alpha=10$ system. (b) $\langle A_{2,far} \rangle$ versus $R$ curves for three systems.  The color here indicates size spans for exponential size distributions [$\alpha = 10, 5,$ and 3; colors matching Fig.~\ref{fig:ndr_radius}(a)] and the open diamonds correspond to the bidisperse system. Solid lines are quadratic fits to guide the eye. The crossing zero point at each solid line is defined as $R^{\ast}$. The inset in (b) shows $R^{\ast}$ obtained from $A_2$ as a function of the polydispersity.}
    \label{fig:far_a2}
    \vspace*{-20pt}
\end{figure}

To verify this assertion, we quantify the behavior of $\Delta \vec{r}_{\rm NA}(x,y) \cdot \hat{r}$ by least squares fitting the field data to $A_2(R_i,r)\cos{2\theta}$, that is, switching from $(x,y)$ to $(r,\theta)$ and taking advantage of the symmetry of Fig.~\ref{fig:ndr_field} to express the magnitude of the flow in terms of the prefactor $A_2(R_i,r)$.  This amplitude varies as a function of distance $r$ to the center of the reference particles. The results for $A_2(R_i,r)$ for several reference droplet radii $R_i$ are shown in Fig.~\ref{fig:far_a2}(a), showing an obvious dependence of $\Delta \vec{r}_{\rm NA}$ on size. For the largest reference particles [${R}_i=5$, the dark purple curve] $A_2$ is negative for all distances $r$ from the reference particles.  This negative $A_2$ indicates that the large particles are strong, and cause the average flow field sketched in Fig.~\ref{fig:sample_example}(b) and quantified in Fig.~\ref{fig:ndr_field}(top right).  The shape of $A_2$ gradually changes with decreasing ${R}_i$. For the smallest reference particles [$R_i=0.5$, the light pink curve], $A_2$ is positive over most of the range, with a small exception at the smallest $r$.  This confirms that these particles are weak, and are the ones whose motion is most often perturbed by the larger particles, quantifying what is seen in Fig.~\ref{fig:ndr_field}. These results are qualitatively different from the bidisperse case where the two $A_2$ curves for the two sizes oscillate, which reflects the pair correlation function and matches the rings visible in the bottom two panels of Fig.~\ref{fig:ndr_field}.

Again, how do these results depend on the size of the reference particles? Here we particularly focus on the far field:  in some cases $A_2>0$ for large $r$ indicating weak particles, and in others, $A_2<0$ indicating strong particles.  We quantify the far field by calculating the average $\langle {A_2}(r) \rangle_r$ over $R_i + 6 \leq r \leq 40$; our results are not sensitive to this choice of averaging region or to, instead, trying to fit the decay of $\langle A_2(r) \rangle$.  The qualitative results discussed above are confirmed in  Fig.~\ref{fig:far_a2}(b): the flow pattern for non-affine motion differs in sign for small reference particles as compared to large reference particles.

These results answer two interesting questions.  First, for a given size distribution, how do we distinguish between ``large'' and ``small'' particles? We propose $\langle A_{2,{\rm far}}(R^*) \rangle = 0$ as the criteria separating the two classes of particles.  For the broadest particle size distribution that we have discussed extensively above, ${R}^* \approx 1.7$. Second, how does this particle size scale depend on the particle size distribution?  The inset to Fig.~\ref{fig:far_a2}(b) shows $R^*$ as a function of the polydispersity $\delta$ of the particle size distributions.  Not surprisingly, $R^*$ grows for broader particle size distributions.  Intriguingly, in our systems the relative fraction of particles with $R_i > R^*$ decreases from 39\% to only 9\% from our narrowest to broadest size distributions.

In this work we have shown that in a sheared amorphous material with high polydispersity, particle size matters.  Large particles are more likely to move affinely, following the imposed shear flow, as they feel the average motion of all of their neighbors. We term such particles as ``strong'' in the sense that they resist being pushed non-affinely by their neighbors.  The imposed shear flow causes those neighbors to detour around the strong particles, which means the smaller a particle is, the ``weaker'' it is and thus the more its motion is nonaffine. We show that one can quantify this by identifying a transition particle radius, $R^*$, separating the two classes of particles. Furthermore, we see that these effects become increasingly important as the particle size distribution broadens. Intriguingly, we demonstrate this distinction still matters, albeit only slightly, for the canonical bidisperse sample with particle size ratio $1:1.4$.  Nonetheless, the behavior of the highly polydisperse samples is qualitatively distinct from the more homogeneous samples with low polydispersity. Our results may have implications, e.g., for diffusive motion in biological cells, which are highly polydisperse crowded environments \cite{ridgway08}.

A further consequence of our work will be on predicting sites of plasticity in highly polydisperse athermal amorphous materials under shear or particle rearrangements at finite temperature. Current analyses typically focus on the rearrangement statistics of only large particles, or implicitly assume via their definition of plastic activity that the qualitative nature of rearrangements are insensitive to particle size~\cite{manning11,cubuk15,schoenholz16,bapst2020unveiling,boattini2021averaging,ding2014soft,patinet2016connecting}. Our results suggest that if one wishes to look for such soft spots in polydisperse materials, a definition of softness that explicitly depends on particle sizes will be necessary.

\begin{acknowledgments}
This material is based upon work supported by the National Science Foundation under Grant No. (CBET-1804186) (YJ and ERW).
\end{acknowledgments}
 

\end{document}